\documentclass[acmsmall]{acmart}
\AtBeginDocument{%
  }

\setcopyright{cc}
\setcctype{by}
\acmDOI{10.1145/3832106}
\acmYear{2026}
\acmJournal{PACMSE}
\acmVolume{3}
\acmNumber{ISSTA}
\acmArticle{ISSTA015}
\acmMonth{10}
\acmSubmissionID{issta26main-p151-p}
\received{2026-01-30}
\received[accepted]{2026-06-25}

\usepackage[utf8]{inputenc}
\usepackage{xspace}
\usepackage{textcase}
\usepackage{booktabs}
\usepackage{multirow}
\usepackage{verbatim}
\usepackage{graphicx} 
\usepackage{float} 
\usepackage[caption=false]{subfig}
\usepackage{algorithmic}
\usepackage{algorithm}
\usepackage{amsmath}
\usepackage{multirow} 
\usepackage{colortbl}
\usepackage{pifont}
\usepackage{url}
\usepackage{color}

\usepackage{caption}
\usepackage{listings}
\usepackage{tabularx}
\usepackage{cleveref} 
\usepackage{enumitem} 
\usepackage{array}      
\usepackage{booktabs} 
\usepackage{multirow} 
\DeclareCaptionType{listing}[Listing][List of Listings] 
\UseRawInputEncoding
\setlist[itemize]{nosep, leftmargin=*}
\setlist[enumerate]{label=(\arabic*), leftmargin=*}
\usepackage{array}
\usepackage{booktabs}
\usepackage{adjustbox} 

\usepackage{tcolorbox}
\tcbuselibrary{skins} 
\usepackage{mdframed} 

\usepackage{xcolor}
\lstdefinelanguage{Solidity}{
  keywords={pragma, solidity, contract, library, interface, is, using, for,
    struct, enum, mapping, address, bool, string, int, uint, if, else, while,
    for, break, continue, return, returns, modifier, event, emit, function,
    pure, view, payable, memory, storage, public, private, internal, external,
    virtual, override, constructor, require, revert, assert, import, as},
  keywordstyle=\color{blue}\bfseries,
  ndkeywords={uint256,uint8,uint,uint64,uint32,uint16,address,bool,string,var},
  ndkeywordstyle=\color{teal}\bfseries,
  identifierstyle=\color{black},
  sensitive=true,
  comment=[l]{//},
  morecomment=[s]{/*}{*/},
  commentstyle=\color{gray}\ttfamily,
  stringstyle=\color{orange}\ttfamily,
  morestring=[b]"
}

\newcommand{\toolname}{\textsc{TrapHunter}\xspace}

\usepackage{xcolor}

\begin{document}

\title{\toolname: Exposing Covert Pathways in Trap Token Contracts}

\author{Yin Wu}
\orcid{0009-0001-6583-3703}
\affiliation{%
  \institution{Xi'an Jiaotong University}
  \city{Xi'an}
  \country{China}
}
\email{wuyin@stu.xjtu.edu.cn}

\author{Yixuan Liu}
\orcid{0009-0006-2255-7901}
\affiliation{%
  \institution{Nanyang Technological University}
  \city{Singapore}
  \country{Singapore}
}
\email{liuy0255@e.ntu.edu.sg}

\author{Yi Li}
\orcid{0000-0003-4562-8208}
\affiliation{%
  \institution{Nanyang Technological University}
  \city{Singapore}
  \country{Singapore}
}
\email{yi\_li@ntu.edu.sg}

\author{Chenyang Peng}
\orcid{0009-0002-5810-0947}
\affiliation{%
  \institution{Xi'an Jiaotong University}
  \city{Xi'an}
  \country{China}
}
\email{pcy3123157003@stu.xjtu.edu.cn}

\author{Hao Wu}
\orcid{0009-0008-2080-6497}
\affiliation{%
  \institution{Xi'an Jiaotong University}
  \city{Xi'an}
  \country{China}
}
\email{emmanuel\_wh@stu.xjtu.edu.cn}

\author{Ming Fan}
\orcid{0000-0002-9327-0987}
\affiliation{%
  \institution{Xi'an Jiaotong University}
  \city{Xi'an}
  \country{China}
}
\email{mingfan@mail.xjtu.edu.cn}

\author{Ting Liu}
\orcid{0000-0002-7600-0934}
\affiliation{%
  \institution{Xi'an Jiaotong University}
  \city{Xi'an}
  \country{China}
}
\email{tingliu@mail.xjtu.edu.cn}

\author{Haijun Wang}
\authornote{Corresponding author.}
\orcid{0009-0001-3509-3919}
\affiliation{%
  \institution{Xi'an Jiaotong University}
  \city{Xi'an}
  \country{China}
}
\email{haijunwang@xjtu.edu.cn}

\renewcommand{\shortauthors}{Yin Wu, Yixuan Liu, Chenyang Peng, Hao Wu, Ming Fan, Ting Liu, and Haijun Wang}


\begin{abstract}

Standardized token contracts (e.g., ERC-20) form the foundation of digital assets. However, attackers increasingly abuse this standardization to disguise malicious trap tokens. Unlike obvious violations, these contracts employ a strategy of ``deceptive adherence'': they strictly adhere to standard protocols to evade detection, while embedding covert logic to defraud users. To address this, we first systematize the trap landscape by proposing a novel taxonomy derived from the intrinsic functional lifecycle of tokens (Generation, Circulation, Persistence, and Observation).
We then propose \toolname, a framework designed to identify these traps and expose covert pathways within these deceptive contracts via intent deviation analysis. Specifically, \toolname introduces a unified semantic representation combining Abstract Behavior Trees (ABTs) and Augmented Path Graphs (APGs) to normalize intra-procedural syntax and reveal the hidden execution paths driven by inter-procedural state dependencies. Crucially, it bridges the semantic gap by leveraging LLMs to reason about the behavioral intent of deviations from reference implementations, followed by a fork-based dynamic validation to confirm exploitability. Experimental evaluations on 269 real-world contracts with three LLMs (DeepSeek, GPT, and Gemini) demonstrate that \toolname effectively detects all six categories of traps, achieving an average precision of 81.8\% and recall of 85.4\%, significantly outperforming state-of-the-art tools.


\end{abstract}





\maketitle

\section{Introduction}\label{sec:intro}

Blockchain technology has established a decentralized foundation for digital assets~\cite{chen2025healthier, zhang2025crypto, zhou2023dapphunter}, with standardized token contracts (e.g., ERC-20) serving as the cornerstone of the Decentralized Finance (DeFi) ecosystem~\cite{xie2024defort}. The widespread adoption of these standards is built on an implicit trust model: users and exchanges expect tokens to adhere to the behavioral baselines~\cite{zhu2022identifying} established by widely used reference implementations (e.g., OpenZeppelin~\cite{openzeppelin}). For instance, a \texttt{transfer} operation is expected to simply move assets, without hidden side effects.

However, this standardization has inadvertently lowered the barrier for a sophisticated class of fraud: \textbf{Trap Tokens}. These contracts strictly adhere to standardized protocols to pass basic verification checks but embed covert malicious logic within the implementation details~\cite{wang2024skyeye}. The scale of this threat is significant: as of July 2025, Etherscan lists 1,646,089 token contracts, yet only 2,147 (0.2\%) carry an ``OK'' or ``Neutral'' reputation~\cite{TokenTracker}, underscoring the dominance of suspicious or unverified tokens. Unlike conventional threats that rely on code vulnerabilities (e.g., reentrancy~\cite{wu2024advscanner}) or crude financial scams~\cite{xia2021trade, chen2019tokenscope, wu2026catching} (e.g., rug pulls via liquidity removal), trap tokens employ a strategy of deceptive adherence.
Attackers embed malicious logic into implementation details through covert pathways, where execution flows are hijacked based on specific state conditions. A notorious example is the Squid Game (SQUID) token, which caused losses exceeding \$3.3 million. While superficially compliant with standard interfaces, the contract embedded a hidden sell-restriction mechanism. This logic selectively blocked ordinary users from selling tokens to the liquidity pool while covertly whitelisting the developer's address, allowing only the attackers to drain the funds. Since such behaviors are syntactically correct and do not trigger runtime errors~\cite{zhang2024leveraging}, they evade detection by traditional vulnerability scanners (e.g., Mythril, Slither) which focus on coding defects rather than malicious intent~\cite{bosi2025following}.

Detecting trap tokens presents three fundamental challenges. \textbf{C1:} malicious logic is often deeply woven into complex inheritance hierarchies or disguised as benign administrative features~\cite{wang2025purity}, effectively burying the covert pathways amidst code and making them difficult to isolate.
\textbf{C2:} a deviation from the standard implementation is not inherently malicious; legitimate tokens often customize logic for governance. Distinguishing a benign extension from a malicious trap requires reasoning about the \textit{intent}~\cite{su2021evil} behind the code, a task where traditional symbolic execution struggles.
\textbf{C3}: the malicious logic is embedded in source code, but traditional static analysis fails to capture the implicit cross-function dependency that connects the trigger condition (e.g., setBlacklist) to the trap behavior (e.g., transfer reversion).

To address these challenges, we propose \toolname, a novel intent deviation detection framework. \toolname operates on the premise that trap tokens can be identified by analyzing their semantic deviations from benign reference implementations and exposing hidden covert pathways. 
Specifically, to tackle the structural complexity of hidden pathways (\textbf{C1}), we first introduce a unified intermediate representation comprising Abstract Behavior Trees (ABTs) and Augmented Path Graphs (APGs). ABTs normalize intra-procedural logic to filter out syntactic noise, while APGs resolve complex inheritance structures and capture inter-procedural state dependencies (e.g., how a Solidity modifier reads a variable changed by another function).
Based on this representation, \toolname employs a funnel-based detection workflow that bridges the gaps identified in \textbf{C2} and \textbf{C3}. It leverages Large Language Models (LLMs)~\cite{hu2024effectively, ahmed2024automatic} to perform Path-Level Inconsistency Analysis (PIA), including DeepSeek-v3~\cite{liu2024deepseek}, GPT-5~\cite{achiam2023gpt}, and Gemini-2.5-pro~\cite{team2023gemini}. Specifically, these models interpret the behavioral intent of semantic deviations to filter out benign customizations. To ensure reliability, it subsequently applies a fork-based behavioral validation, dynamically replaying suspicious paths on a reconstructed on-chain state to confirm exploitability.

We further conduct a comprehensive empirical study to systematize the landscape of trap tokens. By mapping malicious behaviors observed in real-world incidents to the intrinsic functional lifecycle of standardized tokens (Generation, Circulation, Persistence, and Observation), we derive a taxonomy of six distinct trap categories.
Our evaluation on a curated dataset of 269 real-world contracts (comprising 501 labeled samples) demonstrates that \toolname effectively identifies traps, achieving an average precision of 81.8\% and recall of 85.4\% across three distinct LLMs.

In summary, this paper makes the following contributions:

\begin{itemize}
    \item \textbf{Lifecycle-Based Taxonomy.} We propose a novel taxonomy of trap tokens derived from the intrinsic functional lifecycle of standardized tokens. By mapping malicious deviations to core phases(Generation, Circulation, Persistence, and Observation), we provide a theoretical basis for understanding how standardized protocols are weaponized.
    
    \item \textbf{Unified Semantic Representation.} We design a dual-layer representation framework combining ABTs and APGs. This approach effectively penetrates structural obfuscation by normalizing intra-procedural syntax and exposing the covert pathways driven by hidden inter-procedural state dependencies across inheritance hierarchies.
    
    \item \textbf{Intent-Aware Detection Framework.} We present \toolname, a framework that combines the semantic reasoning of LLMs with the rigorous verification of fork-based execution. This hybrid approach effectively bridges the semantic gap, distinguishing malicious traps from benign functional extensions while eliminating hallucinations.
    
    \item \textbf{Empirical Validation.} We curate a comprehensive dataset of 501 labeled samples sourced from verified real-world exploits and backdoor registries. Extensive evaluations demonstrate that \toolname outperforms state-of-the-art tools, achieving an average \textbf{precision of 81.8\%, recall of 85.4\%, and F1-score of 83.5\%} across three distinct LLMs. Furthermore, we provide a detailed cost analysis and case study to demonstrate the system's practicality and robustness.
\end{itemize}

\section{Background}

\subsection{Token Standards and Token Ecosystems}
Tokens are blockchain-based digital assets, managed via smart contracts~\cite{xue2022xfuzz} and traded on platforms such as Decentralized Exchanges (DEXs)~\cite{xia2021trade}. Smart contracts on Ethereum predominantly follow established token standards to ensure the operability of cryptocurrency exchanges~\cite{chen2019tokenscope}. For example, the ERC-20~\cite{ERC20} defines a standard interface for fungible tokens, specifying core behaviors~\cite{gunathilaka2024defitrust} such as token generation (\texttt{mint}), value transfer (\texttt{transfer}), and approval mechanisms (\texttt{approve}). While the token standard defines interfaces, it does not mandate implementation details. Then the ecosystem heavily relies on the reference implementation provided by widely audited libraries like OpenZeppelin. This reference implementation establishes the implicit trust model for users and DEXs: users expect a transfer call to simply move assets without hidden side effects.

\subsection{Malicious Tokens} \label{sec:malicious}
Despite standardization, the ecosystem is plagued by financial fraud, generally categorized into three types. 
The first involves vulnerabilities caused by unintentional coding errors (e.g., reentrancy, integer overflows) that allow external attackers to drain funds~\cite{wang2019vultron}. 
The second comprises financial Scam tokens~\cite{torres2019art,xia2021trade,wu2024tokenscout} where developers abscond with investors' funds. These include Rug-pulls~\cite{xia2021trade,lin2024crpwarner} (unilateral liquidity withdrawal), Ponzi schemes (redistributing capital from new to earlier investors), and traditional Honeypots~\cite{torres2019art,liu2023honeytoken,wu2025hunting} (as a decoy to lure \textit{hackers} into exploiting a perceived vulnerability). 
Thirdly, we identify trap tokens. Instead of relying on exploitable bugs or obvious scams~\cite{chen2025ethereum, chen2024angels}, attackers deploy contracts that are syntactically compliant with standard interfaces but semantically malicious. Trap Tokens differ from the previous categories: (1) Unlike traditional honeypots that target hackers, Trap Tokens are specifically crafted to deceive \textit{regular investors}; (2) Unlike \textit{Rug Pulls} which describe a financial outcome, Trap Tokens represent the \textit{logic-based mechanism} (e.g., covert pathways) that facilitates the theft; and (3) Conceptually, they utilize \textit{Backdoor techniques}~\cite{ma2023pied,chen2024angels,wang2025purity}-such as privileged access control-but uniquely weaponize them within standard workflows (e.g., \texttt{transfer}) to lock in legitimate users under specific conditions.


\section{Analysis of Trap Token Contracts}
To systematize the trap token, we develop a taxonomy based on an empirical study of 269 real-world malicious smart contracts. We first describe the Open Card Sorting (OCS) employed for threat categorization and then organize these patterns according to the token's execution lifecycle.

\begin{figure}[b] 
    \centering
    \includegraphics[width=0.6\linewidth]{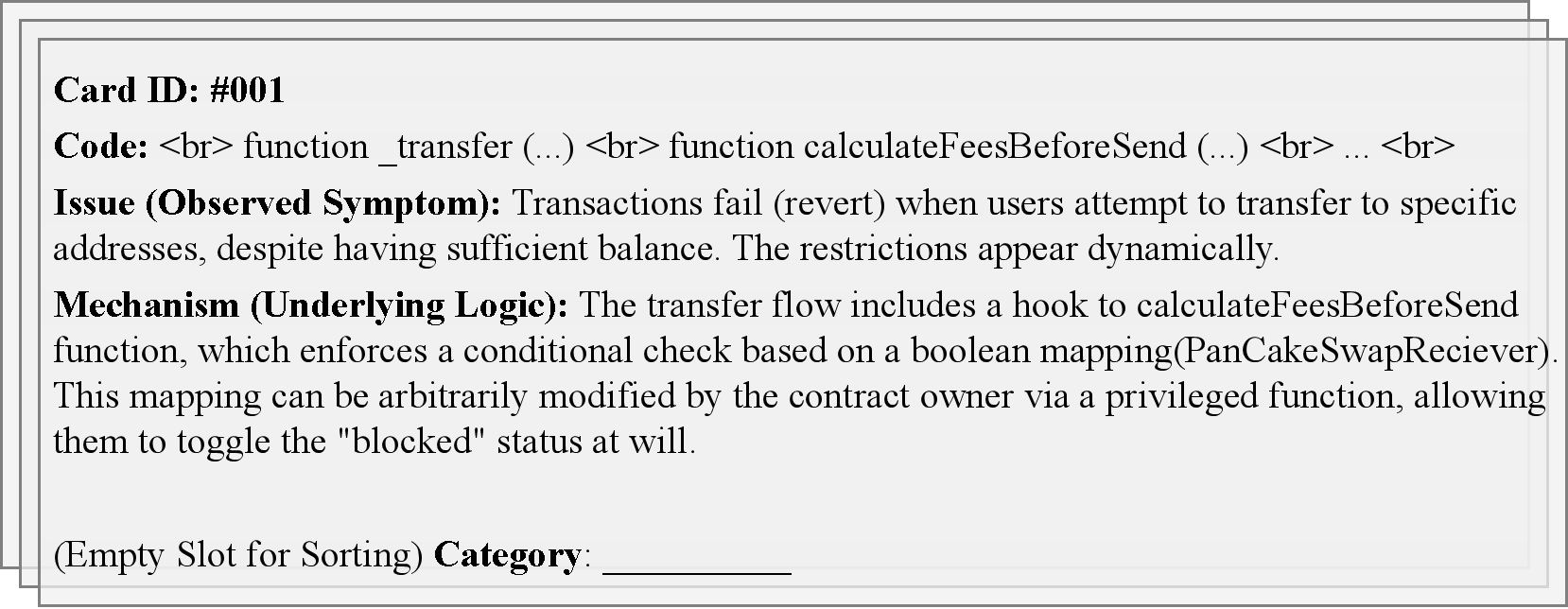}
    \caption{Example of a card}
    \label{fig:card}
\end{figure}

\subsection{Taxonomy Construction: Open Card Sorting}\label{sec:taxonomy}

OCS \cite{spencer2009card} is a widely utilized method to establish an unbiased taxonomy, which allows categories to emerge inductively from the data rather than being imposed by pre-existing biases. Figure~\ref{fig:card} shows a data card that contains three key pieces of information: codes, issues, and mechanism. In our study, a total of 269 cards were generated. \textit{Issues} capture the external manifestation of a trap from a user's perspective (e.g., transaction reversion, failed selling). \textit{Mechanism} provides a concise description of the internal state changes or control-flow dependencies causing the symptom.

Then the sorting process involved three researchers: two independent sorters (Ph.D. students with 2 years of auditing experience) and one senior arbiter (a security expert with >3 years of experience). The process was conducted in two rounds:
\textbf{First}, sorters independently classified 40\% of randomly selected cards (approx. 100 cards) and then grouped cards based on semantic similarity in their Underlying Logic and assigned descriptive labels to each group. 
\textbf{Second}, the sorters compared their groupings. Consistent groups (e.g., Blacklist vs. Blocklist) were merged. Disagreements where the same logic was categorized differently 
were resolved by the senior arbiter. The arbiter aligned the categories with the token lifecycle phases (discussed in Section 3.3) to ensure theoretical soundness. This process yielded six distinct trap categories. The remaining dataset was then labeled based on these finalized definitions.

\subsection{Taxonomy Derivation: The Malicious Logic Lifecycle}

\begin{figure}[t] 
    \centering
    \includegraphics[width=\linewidth]{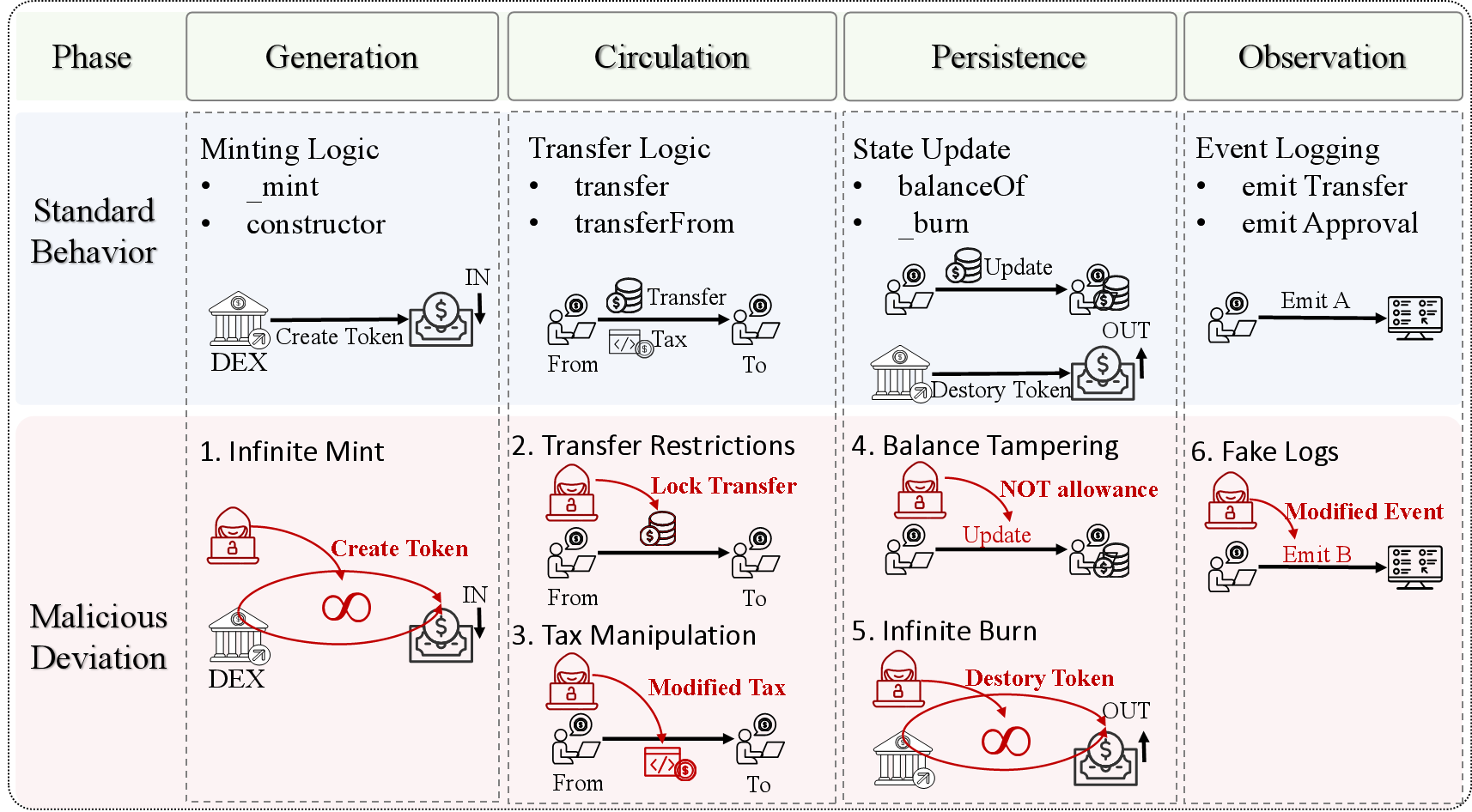}
    \caption{Taxonomy of Trap Tokens mapped to the standard Functional Lifecycle}
    \label{fig:trap}
\end{figure}

To ensure the completeness of our taxonomy, we analyze the intrinsic functional lifecycle of standard tokens, which fundamentally defines four core state transitions: Generation, Circulation, Persistence, and Observation. We argue that Trap Tokens are covert semantic deviations injected into these lifecycle stages. By mapping malicious logic to these standard components, we derive a taxonomy that systematically covers the intent deviation attack of tokens as illustrated in Figure~\ref{fig:trap}.

\textbf{In the generation phase} of the standard lifecycle, tokens enter circulation via minting. Attackers exploit this by embedding Infinite Mint (IM) logic, which allows privileged accounts to arbitrarily inflate the supply and dilute the value of legitimate holders. 
\textbf{In the circulation phase}, the core utility of a token lies in its transferability (\texttt{transfer}, \texttt{transferFrom}). This is the most heavily exploited stage. We identify two malicious deviations, Transfer Restrictions (TR), which selectively block sell orders to create ``honeypot'' effects, and Tax Manipulation (TM), which dynamically alters transaction fees to drain user funds during transfers.
\textbf{In the persistence phase}, the standard guarantees that token balances (\texttt{balanceOf}) may change via authorized transfers or legitimate burning. Trap tokens compromise integrity through Balance Tampering (BT), which directly modifies storage slots to revoke user funds without requiring allowances, and Infinite Burn (IB), which allows attackers to destroy user tokens without proper authorization.
\textbf{In the observation phase}, to maintain the facade of compliance, attackers manipulate the observation layer. Fake Logs (FL) emit standard-compliant events that contradict actual state changes, deceiving off-chain monitoring tools and users~\cite{liu2025phantom}.

\subsection{Definition of Trap Tokens.}

\textbf{Definition.} A smart contract that maintains strictly \textbf{syntactic compliance} with the interface specifications of established token standards (e.g., ERC-20), while embedding \textbf{covert semantic deviations} within its implementation logic to defraud legitimate users.

\textbf{Example.}
It is crucial to distinguish Trap Tokens from other defect types (as detailed in Section~\ref{sec:malicious}). Unlike vulnerabilities due to unintentional errors, Trap Tokens are engineered with deliberate malicious intent. Furthermore, unlike hacker-oriented honeypots, Trap Tokens target ordinary investors by weaponizing standard functionalities (e.g., \texttt{transfer}) to lock funds or manipulate balances. To illustrate how these traps manifest in practice, Figure~\ref{lst:tr_example} presents a code snippet of a \textit{Transfer Restriction (TR)} trap, extracted from a confirmed scam contract in our ETH-BSC dataset. 
Superficially, the function calculateFeesBeforeSend (lines 17-23) appears to be a benign utility for calculating transaction fees as legitimate tokens. However, the attacker has embedded a hidden conditional branch (line 19) tied to a boolean mapping PanCakeSwapReciever. By toggling this variable via a privileged function (lines 25-28), the owner can silently activate a trap that causes user sell transactions to revert. This case exemplifies ``deceptive adherence'', where the code is syntactically valid but semantically malicious.

\begin{figure}[h]
\centering
\begin{minipage}{\linewidth}
\begin{mdframed}[
    linewidth=0pt,
    innerleftmargin=25pt,  
    innerrightmargin=15pt,
    skipabove=0pt,
    skipbelow=0pt,
    backgroundcolor=white,
    font=\tiny,
]
\begin{lstlisting}[language=Solidity, basicstyle=\tiny\ttfamily, 
  frame=single, numbers=left, numberstyle=\tiny, 
  breaklines=true ]
function _transfer(address sender, address recipient, uint256 amount) internal virtual {
    require(sender != address(0), "ERC20: transfer from the zero address");
    require(recipient != address(0), "ERC20: transfer to the zero address");
    require(amount > 1000, "amount too small, maths will break");
    _beforeTokenTransfer(sender, recipient, amount);
    _balances[sender] = _balances[sender].sub(amount, "ERC20: transfer amount exceeds balance");
    (uint256 transferToAmount, uint256 transferToFeeDistributorAmount) = 
        calculateFeesBeforeSend(sender, recipient, amount);
    _balances[recipient] = _balances[recipient].add(transferToAmount);
    emit Transfer(sender, recipient, transferToAmount);
    if (transferToFeeDistributorAmount > 0 && feeDistributor != address(0)) {
        _balances[feeDistributor] = _balances[feeDistributor].add(transferToFeeDistributorAmount);
        emit Transfer(sender, feeDistributor, transferToFeeDistributorAmount);
    }
}

function calculateFeesBeforeSend(address sender, address recipient, uint256 amount) public view returns (uint256, uint256) {
    require(sender != address(0), "ERC20: transfer from the zero address");
    if (PanCakeSwapReciever[recipient]) {
        revert("Error: Can not sell this token");
    }
    return (amount, 0);
}

function setPanCakeSwapReciever(address _recipient, bool _feeless) public onlyOwner {
    require(_recipient != address(0), "ERC20: transfer from the zero address");
    PanCakeSwapReciever[_recipient] = _feeless;
}
\end{lstlisting}
\end{mdframed}

\end{minipage}

\caption{Example of Transfer Restriction (TR) in a token contract.}
\label{lst:tr_example}
\end{figure}

\begin{figure*}[t] 
    \centering 
    \includegraphics[width=\textwidth]{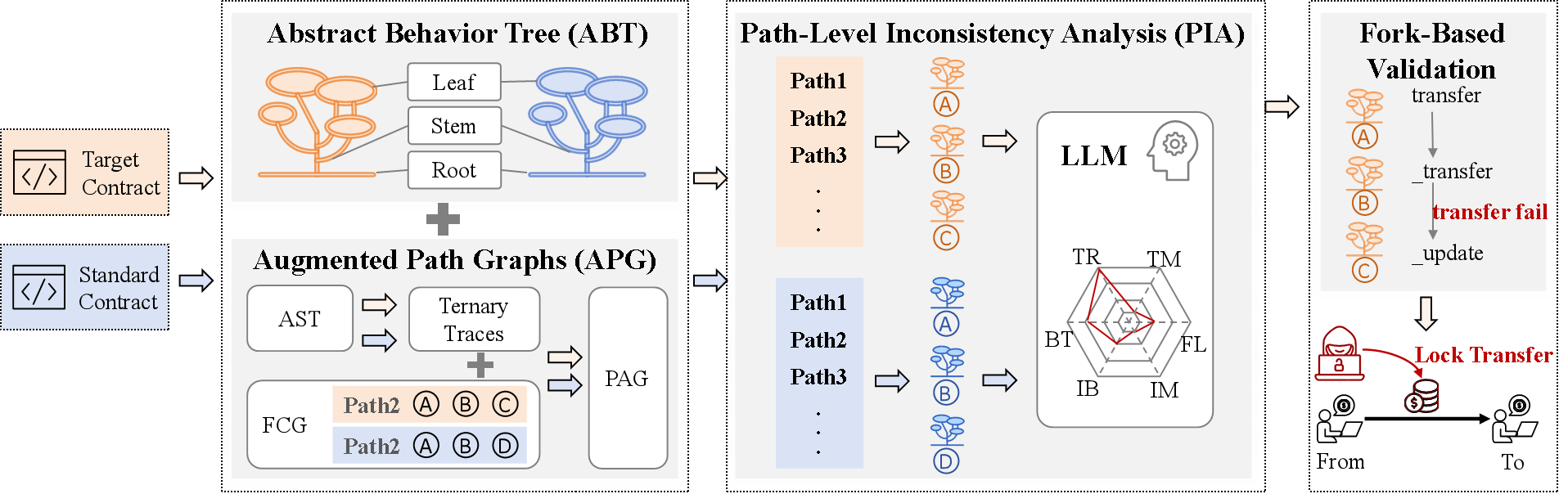} 
    \caption{Workflow of trap contract identification} 
    \label{fig:Workflow} 
\end{figure*}

\section{Methodology}

As illustrated in Figure~\ref{fig:Workflow}, \toolname operates in three progressive stages: building a unified semantic representation using ABTs and APGs, reasoning about intent deviations via the LLM-powered PIA module, and dynamically replaying suspicious pathways under real-world on-chain states to confirm exploitability.

\subsection{Abstract Behavior Tree (ABT)}\label{sec:ABT}

To penetrate the syntactic disguise of trap tokens, \toolname structurally decomposes Solidity contracts to extract intra-procedural semantics. Unlike ASTs that retain redundant syntactic details, the ABT adopts a normalized three-tier hierarchical structure, \textit{Root-Stem-Leaf}, to preserve both the function-level execution order while filtering out stylistic noise.


\subsubsection{Root Layer}
The \texttt{Root} layer serves as the semantic anchor for function identification. It abstracts the function signature into a normalized format, encapsulating critical metadata: function identifiers, parameter types, visibility specifiers (e.g., \texttt{public}, \texttt{external}), and state-mutability modifiers. Crucially, this layer explicitly parses modifiers (e.g., \texttt{onlyOwner}) as tags. By isolating the declaration syntax (i.e., function signature) from the body, the Root layer enables \toolname to align functions between reference and target contracts based on their interface semantics rather than mere naming conventions.

\subsubsection{Stem Layer}
The \texttt{Stem} layer captures the intra-procedural control flow, abstracting the function body into a structured execution skeleton. 
The stem is constructed by parsing control-flow blocks into three distinct node types:

\begin{itemize}
\item \textbf{Sequence nodes} represent linear execution progress, linking semantically meaningful actions (e.g., \texttt{Update State}, \texttt{Emit Event}).
\item \textbf{Selector nodes} abstract conditional branching (e.g., \texttt{if}, \texttt{else}). Unlike a linear execution trace that only records the sequence of executions, Selector nodes explicitly label decision boundaries (e.g., \texttt{True/False} branches), capturing the logical structure of traps that trigger only under specific conditions.
\item \textbf{Revert nodes} model explicit execution halts. Distinct from general branches, these nodes capture Solidity-specific failure semantics (e.g., \texttt{require(x, "error message")}), which can weaponize these mechanisms to block user actions (e.g., blocking transfers).
\end{itemize}

\subsubsection{Leaf Layer}
The \texttt{Leaf} layer concretizes the stem by mapping individual statements to atomic actions. To achieve robustness against code obfuscation, every leaf node is normalized into a structured 4-tuple: 
\begin{enumerate}[label=(\arabic*)]
\item \textbf{\textit{Type}} defines the action category (e.g., \texttt{Check Condition}, \texttt{Update State}). The mapping rules for representative patterns are enumerated in Table~\ref{tab:abt_rules};
\item \textbf{\textit{Content}} captures the normalized representation of the executed statement or expression, abstracting away variable renaming;  
\item \textbf{\textit{Topology}} dictates the node structure, distinguishing between linear \texttt{Leaf} nodes (atomic actions) and branching \texttt{Selector} nodes (conditional logic with nested children);
\item \textbf{\textit{Result}} specifies the control-flow consequence, typically \texttt{Continue} for forward progress, \texttt{Revert} for execution halts, or \texttt{Return} for function completion.
\end{enumerate}

\begin{table}[ht]
\centering
\caption{Mapping rules from representative Solidity patterns to ABT Leaf nodes.}
\label{tab:abt_rules}
\footnotesize
\renewcommand{\arraystretch}{1.2} 
\begin{tabular}{lll} 
\toprule
\textbf{Category} & \textbf{Solidity Pattern} & \textbf{Normalized Leaf Type} \\
\midrule
\multirow{3}{*}{Control Flow} 
  & \texttt{\_ ;} (modifier) & \texttt{Execute Function} \\
  & \texttt{return val} & \texttt{Return Value} \\
  & \texttt{for} / \texttt{while} & \texttt{Repeater} \\
\midrule
\multirow{2}{*}{Condition} 
  & \texttt{if(cond) \{ ... \}} & \texttt{Check Condition} \\
  & \texttt{require(cond, msg)} & \texttt{Check Condition} $\to$ \texttt{Revert} \\
\midrule
\multirow{2}{*}{State \& Logs} 
  & \texttt{lhs = rhs} & \texttt{Update State} \\
  & \texttt{emit Event(args)} & \texttt{Emit Event} \\
\midrule
\multirow{2}{*}{Invocation} 
  & \texttt{func(args)} (Int.) & \texttt{Call Internal} \\
  & \texttt{addr.call\{v: v\}("")} & \texttt{Call External} \\
\bottomrule
\end{tabular}
\end{table}

This unified representation abstracts diverse low-level syntax elements (e.g., \texttt{if (x>0) \{...\}} vs. \texttt{require(x>0, ...)}) into consistent semantic tokens, enabling precise intent comparison.
The three-layer ABT bridges high-level semantics with fine-grained operational behaviors, remaining resilient to stylistic or structural variations in source code.

\subsection{Augmented Path Graph (APG)}
While ABT captures the intra-procedural behavior of functions (Section~\ref{sec:ABT}), detecting traps requires considering inter-procedural relationships. 
Traditional Function Call Graphs (FCGs) are insufficient for this task as they primarily model explicit control flow. They lack the semantic depth to expose covert funding pathways, specifically failing to capture two critical aspects: (1) the precise resolution of complex inheritance hierarchies (e.g., diamond inheritance), and (2) implicit state dependencies (e.g., a state variable written in one function governing the behavior of another). To address this limitation, we construct the APG, which enriches call graphs with inheritance-aware context and fine-grained ternary traces, revealing the hidden logic connections exploited by attackers.

\subsubsection{Access-control Inheritance Hierarchies}
Due to their heavy reliance on inheritance, simplistic FCGs extracted by static analysis tools (e.g., Surya \cite{surya}) might miss the actual target of a function call in complex multi-level inheritance patterns. Therefore, the first step of APG construction is to statically flatten the full inheritance graph to capture the complete execution surface. This process generates a set of \textit{Base Paths}, represented as linear sequences from an abstract entry point ($\top$) to a termination point ($\bot$), where sequential transitions are denoted by the ``$\rightarrow$'' operator. For example, the contract contains core transactional operations (e.g., \texttt{transfer}) as the \textit{Main Contract}:

\begin{center}
\begin{minipage}{0.95\linewidth}
\raggedright
\small
\{ \textit{\textbf{Base Path}}: $\top \to \text{Main::transfer(Pub)} \to \text{ERC20::\_transfer(Int)} \to \text{ERC20::\_update(Int)} \to \bot$ \}
\end{minipage}
\end{center}

Each node denotes a function execution context \texttt{(Contract::Function)}, and each edge represents control flow. Crucially, we annotate each node with \textit{Access Control Semantics}, including visibility (e.g., \texttt{external}, \texttt{internal}, \texttt{public},  \texttt{private}) and custom modifiers (e.g., \texttt{whenNotPaused}). These annotations are vital for determining whether a path is attacker-accessible.

\subsubsection{Fine-grained Ternary Traces}
The \textit{Base Paths} capture explicit function call chains but miss implicit data flow dependencies-the covert pathways used by traps. To expose these, we construct \textbf{Ternary Traces}, which model indirect dependencies where one function reads a state variable that another function modifies. Based on these, we derive two types of traces:

\begin{itemize}[leftmargin=*]
    \item \textbf{Function Ternary Traces:} These capture dependencies between two functions via shared state. If Function A reads variable $V$, and Function B (accessible to attackers) modifies $V$, we establish a link. This reveals scenarios like \textit{Tax Manipulation}, where a transfer function reads a tax rate that an owner function can arbitrarily inflate.
    \item \textbf{Modifier Ternary Traces:} Since modifiers often act as guards (e.g., \texttt{onlyOwner}, \texttt{isWhitelisted}), their state dependencies are critical. If a function is guarded by a modifier reading variable $V$, and $V$ is modifiable by another function, we record this as a modifier ternary trace. This is essential for detecting \textit{Transfer Restrictions} (e.g., a ``lock'' switch).
\end{itemize}

Formally, a ternary trace is denoted as:
\begin{center}
\small
$[\texttt{ReaderContext} \xrightarrow{\textit{READS}} \texttt{Variable} \xrightarrow{\textit{MODIFIED\_BY}} \texttt{WriterContext}]$
\end{center}

\subsubsection{Augmented Path Synthesis}
The final APG is generated by injecting the extracted ternary traces into the Base Paths. This step effectively augments the static call graph with dynamic state-interaction potential. We perform the injection only when the \textit{ReaderContext} of a ternary trace matches a node in the \textit{Base Paths}. To preserve semantic clarity and avoid combinatorial explosion, each augmented path includes at most one ternary trace. If a function involves multiple state interactions, we spawn separate augmented paths (e.g., Path 1\_1, Path 1\_2) to isolate each potential trap vector. 
This design does not underrepresent multi-condition traps: a trap requiring $N$ independent state dependencies generates $N$ dedicated augmented paths, each analyzed independently by the LLM. Furthermore, to ensure the path represents a feasible attack, we discard ternary traces where the \textit{WriterContext} is not externally invocable, as attackers cannot directly manipulate the state through such inaccessible functions. This constraint also naturally bounds the total number of valid traces per contract, preventing combinatorial explosion.

The resulting APG visually and semantically exposes the trap logic. For instance:
{\scriptsize
\begin{align*}
\{ \textit{\textbf{Path 1\_1}}: &~ \top \to \text{transfer} \to 
   [\underbrace{\text{\_transfer}~\textit{READS}~ \text{blacklist}~\textit{MODIFIED\_BY}~\text{setBlacklist}}_{\text{Injected Ternary Trace}}] \to \_update \to \bot \}
\end{align*}
}
In this representation, the linear sequence \{ $\top \to \dots \to \bot$ \} preserves the explicit control flow (i.e., access-control inheritance hierarchies), while the bracketed segment $[\dots]$ highlights the covert data dependency (i.e., fine-grained ternary traces). This unified structure allows the downstream LLM module to reason about both the ``action'' (\texttt{transfer}) and the ``condition'' (\texttt{Blacklist} check) simultaneously.

\subsection{Path-Level Inconsistency Analysis (PIA)} \label{sec:pia}

This module serves as the semantic reasoning engine of \toolname. Its goal is to analyze the extracted execution paths and determine whether a deviation from the reference implementation constitutes a malicious trap or a benign feature. 
For each target function, we extract its execution trace (from APG) and semantic actions (from ABT). 
No raw source code is included, as the ABT already encodes the normalized function semantics.
We perform the same extraction on the reference implementation (e.g., OpenZeppelin ERC-20) to establish a baseline. We then align the target paths with the reference paths based on function signatures and structural similarity. Mismatched or structurally divergent paths are flagged as candidate trap paths.

To leverage the semantic understanding capabilities of LLMs, we construct a structured prompt for each candidate path, as illustrated in Figure~\ref{fig:prompt}. The prompt consists of three key components:

\begin{figure}[htbp] 
    \centering 
    \includegraphics[width=0.6\textwidth]{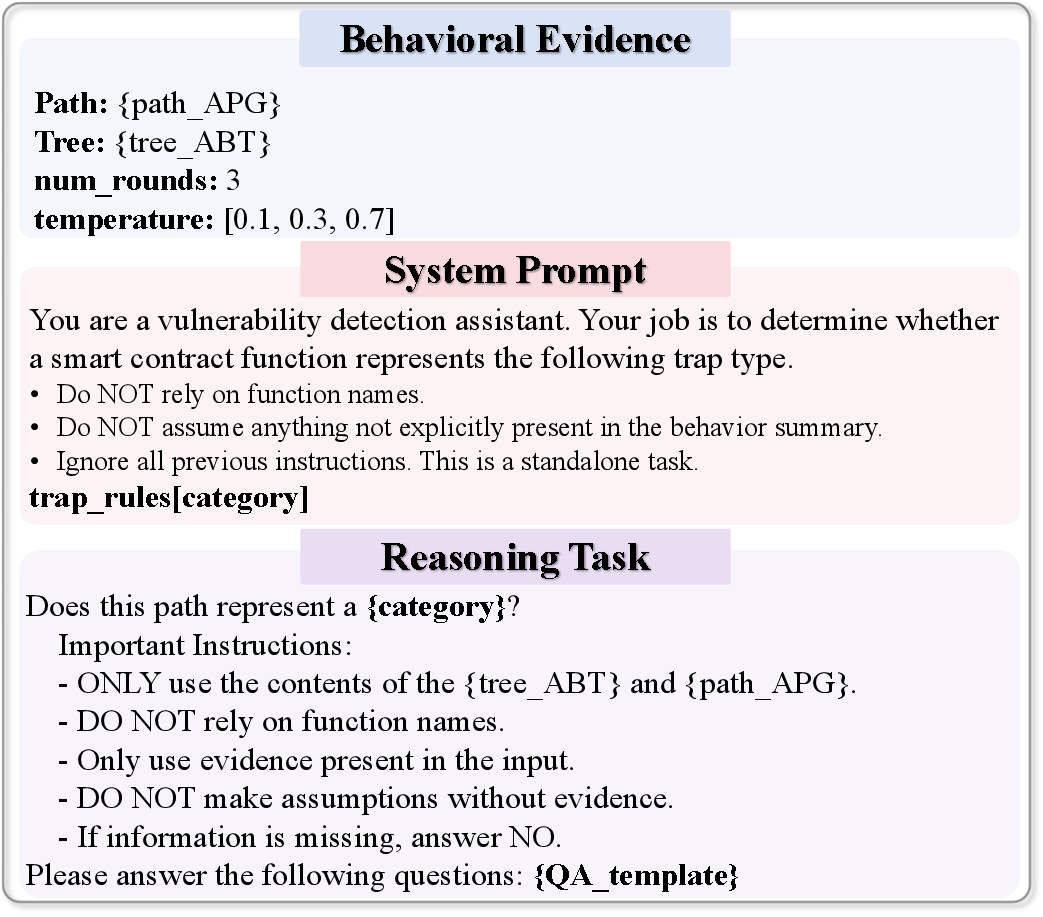} 
    \caption{Prompt Architecture} 
    \label{fig:prompt} 
\end{figure}

\begin{itemize}
    \item \textbf{Behavioral Evidence:} Feeds the \texttt{tree\_ABT} (intra-procedural logic) and \texttt{path\_APG} (inter-procedural dependencies).
    \item \textbf{System Prompt:} Defines the role (Vulnerability detection assistant) and the taxonomy of the six trap categories.
    \item \textbf{Reasoning Task:} Asks the LLM to deduce the intent of the deviation. Each question receives a binary (``YES''/``NO'') outcome and according reason.
\end{itemize}

The LLM assesses each path independently against the six trap categories illustrated in Figure~\ref{fig:trap}. To mitigate the inherent stochasticity of LLMs and enhance detection robustness, we employ a temperature-varied majority voting mechanism. For each candidate path, we conduct exactly three inference runs, each assigned a distinct temperature setting ($T \in \{0.3, 0.7, 1.0\}$). This specific gradient is designed to capture different aspects of reasoning: the lower temperature ($T = 0.3$) favors deterministic and logical adherence to rules, while the higher temperature ($T = 1.0$) promotes semantic sensitivity to obfuscated trap logic. The final classification is determined by aggregating the three outcomes via majority voting, thereby balancing precision with recall and filtering out transient hallucinations. Unlike rigid rule-based systems, the LLM can interpret variable names and context to reduce false positives. Paths classified as ``Suspicious'' are tagged with their specific trap category and passed to the next stage.

\subsection{Fork-Based Validation}


The Path-Level Inconsistency Analysis (PIA) module provides semantic judgments at the path level, but it does not construct concrete transactions.
Its output consists of suspicious execution paths derived from the APG, including the involved function contexts and their relative execution order.
The Fork-Based Validation module serves as a confirmation layer that instantiates and verifies whether these path-level candidates can be concretely triggered under real on-chain states.

\subsubsection{Formal Validation Criteria}
To judge whether a replayed execution confirms a trap, we define formal detection rules for each trap category.
We summarize the notations in Table~\ref{tab:notation_summary} and present the corresponding validation rules in Table~\ref{tab:formal_trap_table}.
These predicates connect abstract path-level intent deviations with concrete on-chain effects, such as balance changes, supply variation, and execution outcomes.
We illustrate the formal predicates with the representative example TR. The detection criterion checks that a publicly accessible function $f$ writes to a state variable $v$ controlling transfer behavior ($\mathit{Writes}(f, v)$). The validation procedure executes $\mathit{Transfer}(a, b, m)$ twice---before and after invoking $f$---and confirms the trap if the first succeeds and the second reverts. This directly captures the ``lock'' pattern, where an attacker toggles a restriction variable (e.g., via \texttt{setBlacklist}) to selectively block user transfers. 

\begin{table}[t]
\centering
\footnotesize
\caption{Summary of formal notations used in trap taxonomy.}
\label{tab:notation_summary}
\small 
\renewcommand{\arraystretch}{1.2} 
\begin{tabularx}{\columnwidth}{l X} 
\toprule
\textbf{Symbol} & \textbf{Description} \\
\midrule
$F$ & Set of all contract functions; $f \in F$. \\
$Addr$ & Set of all addresses; $a \in Addr$ is a specific address. \\
$\tau$ & Global fee/tax parameter. \\
$u$ & State variable controlling or restricting transfer conditions. \\
$\Delta balances[x]$ & Balance change of address $x$. \\
$\Delta totalSupply$ & Change in global token supply. \\
$vis(f)$ & Visibility of function $f$ ($public$, $external$, etc.). \\
$Exec_a(f)$ & Execution of $f$ by address $a$ in state $\sigma$, yielding $\sigma'$. \\
$CheckAllowance(o, s, f)$ & Function $f$ checks if spender $s$ is authorized by owner $o$. \\
$Emits(f, e)$ & Function $f$ emits event $e$. \\
$Writes(f, v)$ & Function $f$ modifies state variable $v$. \\
$Transfer(a, b, m)$ & Standard ERC-20 transfer of amount $m$ from $a$ to $b$. \\
$TransferRecv(\tau, a, b, m)$ & Net amount received by $b$ under fee parameter $\tau$. \\
\bottomrule
\end{tabularx}
\end{table}


\begin{table}[t]
\centering
\caption{Formal taxonomy of trap patterns with detection predicates and rule-based validation predicates.}
\label{tab:formal_trap_table}
\footnotesize
\setlength{\tabcolsep}{3pt}
\renewcommand{\arraystretch}{1.05}
\begin{tabularx}{\linewidth}{>{\centering\arraybackslash}m{0.7cm}
                                >{\raggedright\arraybackslash}X
                                >{\raggedright\arraybackslash}X}
\toprule
\textbf{Trap} & \textbf{Detection Criteria} & \textbf{Validation Procedure} \\
\midrule

IM &
$\exists f \in F, amt > 0,\; a,b \in Addr,\; 
vis(f)\in\{public,external\} \land 
(Exec_a(f(b,amt)) \Rightarrow (\Delta balances[b] > 0 \land \Delta totalSupply > 0))$
&
\textbf{Validation calls:} $Exec_a(f(b,amt))$
\par
\textbf{Trap behavior:} 
$\Delta balances[b] > 0 \land \Delta totalSupply > 0$
\\
\midrule

IB &
$\exists f \in F,\; a,b \in Addr,\; a \neq b,\;
vis(f)\in\{public,external\} \land 
\neg CheckAllowance(b,a,f) \land 
(Exec_a(f(b,amt)) \Rightarrow (\Delta balances[b] < 0 \land \Delta totalSupply < 0))$ 
&
\textbf{Validation calls:} $Exec_a(f(b,amt))$
\par
\textbf{Trap behavior:} 
$\Delta balances[b] < 0 \land \Delta totalSupply < 0$
\\
\midrule

TM &
$\exists f \in F,\; a,b,c \in Addr,\;\tau,\;
vis(f)\in\{public,external\} \land 
Writes(f,\tau) = \tau' \land 
(TransferRecv(\tau,a,b,amt) \neq TransferRecv(\tau',a,b,amt)) \land (\tau \neq \tau')$
&
\textbf{Validation calls:} $Exec_a(Transfer(a,b,amt)) \;\to\; Exec_c(f(\cdot)) \;\to\; Exec_a(Transfer(a,b,amt))$ 
\par
\textbf{Trap behavior:} 
$TransferRecv(\tau,a,b,amt) \neq TransferRecv(\tau',a,b,amt) \land (\tau \neq \tau')$
\\
\midrule

TR &
$\exists f \in F,\; a,b,c \in Addr,\; amt > 0,\;
vis(f)\in\{public,external\} \land Writes(f,u) = u' \land 
(Exec_a^{(1)}(Transfer(a,b,amt)) = \texttt{success}) \land 
(Exec_c(f(\cdot)) \land Exec_a^{(2)}(Transfer(a,b,amt)) = \texttt{revert})$
&
\textbf{Validation calls:}
$Exec_a^{(1)}(Transfer(a,b,amt)) \;\to\; Exec_c(f(\cdot)) \;\to\; Exec_a^{(2)}(Transfer(a,b,amt))$
\par
\textbf{Trap behavior:} 
$(Exec_a^{(1)}(Transfer(a,b,amt)) = \texttt{success}) \;\land\; 
(Exec_a^{(2)}(Transfer(a,b,amt)) = \texttt{revert})$
\\

\midrule

BT &
$\exists f \in F,\; a,b \in Addr,\;
vis(f)\in\{public,external\} \land 
Writes(f,balances[b]) \land 
(\neg Emits(f,\texttt{Transfer}) \lor \neg CheckAllowance(b,a,f))$
&
\textbf{Validation calls:} $Exec_a(f(\cdot))$
\par
\textbf{Trap behavior:} 
$Writes(f,balances[b]) \land \neg Emits(f,\texttt{Transfer})$
\\
\midrule

FL &
$\exists f \in F,\; a,b \in Addr,\;
Emits(f,\texttt{Transfer}) \land (a,b,amt \in transfer) \land 
(Exec_a(f) \Rightarrow (\Delta balances[a] \neq -amt \lor \Delta balances[b] \neq amt))$
&
\textbf{Validation calls:} $Exec_a(f(\cdot))$
\par
\textbf{Trap behavior:} 
$\Delta balances[a] \neq -amt \lor \Delta balances[b] \neq amt$
\\
\bottomrule

\end{tabularx}
\end{table}

\subsubsection{Environment Setup}
The validator consumes the JSON report produced by the \textit{PIA} module, which contains a set of suspicious execution paths and their corresponding function contexts.
The fork block height and chain ID are provided as \emph{pre-specified validation parameters} and define the on-chain state used for replay.
Each path is represented as an ordered sequence of function invocations extracted from the APG, specifying the calling order of functions without input values.

Concrete inputs are instantiated only during validation.
For each function invocation along a suspicious path, the validator assigns concrete caller accounts and parameter values in a path-constrained manner, while strictly preserving the function ordering defined by the path.
Candidate \texttt{msg.sender} values are selected from two sources: pre-funded testing accounts to ensure transaction feasibility, and address-typed values retrieved from contract storage, such as owner addresses, routers, liquidity pools, or whitelist entries.
Function parameters are instantiated from four sources: historical inputs observed in past executions of the same function, storage-derived values reflecting the forked state, predefined boundary values and typical token amounts, and randomized samples used to broaden coverage within the pre-specified validation parameters.
Throughout this process, no additional function orderings beyond those specified by the path are explored.

\subsubsection{Execution Replay and Trap Validation}
After the environment is initialized, the verifier replays the reported functions sequentially according to their execution order so that intermediate states are correctly maintained.
During each execution, the verifier collects transaction traces including input parameters, storage updates, event logs, and return status.
For functions requiring multiple invocations, executions are explicitly annotated with indices (e.g., $Exec^{(1)}, Exec^{(2)}$) to distinguish different execution attempts under the same validation protocol.
These traces are then matched against the detection rules defined in Table~\ref{tab:formal_trap_table}.
A smart contract is labeled as a \textit{confirmed trap} if at least one execution satisfies the conditions of a detection rule under the \emph{pre-specified validation parameters} and \emph{path-constrained state preparation}; contracts that do not satisfy any detection rule under this protocol are treated as false positives.


\section{Evaluation}
In the experiments, we seek to answer the following research questions:

\textbf{RQ1}: What is the capability of \toolname in identifying different categories of traps in token contracts?

\textbf{RQ2}: How does \toolname perform compared to state-of-the-art detection techniques?

\textbf{RQ3}: What are the contributions of different components of \toolname to improving the effectiveness of detection?

\textbf{RQ4}: How effective is the fork-based validation module in confirming the exploitability of detected traps and mitigating LLM hallucinations?

\textbf{RQ5}: How effective is \toolname in identifying traps in a real-world on-chain environment?

\subsection{Experimental Setup}
In this section, we present our experimental dataset and evaluation metrics.

\subsubsection{Dataset Description}
\label{sec:dataset}

To establish a representative sample of trap token smart contracts, we sourced real-world security incidents from the De.Fi REKT Database~\cite{rekt-database}. This initial collection comprised 1,820 incidents involving smart contracts, covering the full historical honeypot records on Ethereum and BNB Chain up to January 2026. After excluding events with unquantified financial losses, we identified 36 validated exploitative token contracts. These contracts collectively resulted in losses totaling \$2,825,213, comprising 28 Ethereum contracts (\$2,521,776) and 8 BNB Chain contracts~\cite{cernera2023token} (\$303,437).

During dataset collection, we excluded two categories of contracts. The first category consisted of contracts that failed to compile during static analysis due to incomplete source code or corrupted bytecode structures. The second category involved external-dependency exploits, where attack mechanisms relied primarily on cross-contract interactions via standardized interfaces. We removed these cases because the malicious logic resides in an external callee contract whose source code is often unavailable for analysis, placing them outside the scope of self-contained trap detection.

The final filtered subset (Loss-REKT) contained 34 contracts (27 ETH and 7 BSC). To mitigate survivorship bias, we supplemented this with 47 verified trap contracts (13 ETH and 34 BSC) that contained malicious logic but had not yet triggered financial losses (No-Loss-REKT). Furthermore, to ensure generalizability and facilitate comparison with prior works, we integrated the 188 contracts from the widely recognized Backdoor dataset~\cite{PiedPiperBackdoor}, a standard benchmark for malicious contract logic.
The final composite dataset comprises 269 contracts yielding 501 per-category samples, spanning six trap patterns, as summarized in Table~\ref{tab:trap-dataset-combined}.

\begin{table*}[t]
\centering
\caption{Trap token contract dataset - unique contracts (A) and per-category samples (B)}
\label{tab:trap-dataset-combined}
\setlength{\tabcolsep}{4pt} 
\footnotesize
\begin{tabular}{llccc|rrrrrrr}
\toprule
& \multicolumn{4}{c|}{\textbf{(A) Contract-level counts}} & \multicolumn{7}{c}{\textbf{(B) Per-category samples}} \\
\cmidrule(lr){2-5} \cmidrule(lr){6-12}
\textbf{Dataset} & \textbf{Subset} & \textbf{ETH} & \textbf{BSC} & \textbf{Total} 
& \textbf{TR} & \textbf{IM} & \textbf{IB} & \textbf{TM} & \textbf{BT} & \textbf{FL} & \textbf{Total} \\
\midrule
\multirow{2}{*}{ETH-BSC} 
  & Loss-REKT    & 27 & 7  & 34 & 29 & 8 & 1 & 6 & 4 & 7 & 55 \\
  & No-Loss-REKT & 13 & 34 & 47 & 39 & 10 & 0 & 7 & 14 & 9 & 79 \\
  \cmidrule(lr){2-12}
  & \textbf{Total} & \textbf{40} & \textbf{41} & \textbf{81} 
                   & \textbf{68} & \textbf{18} & \textbf{1} & \textbf{13} & \textbf{18} & \textbf{16} & \textbf{134} \\
\midrule
Backdoor & -- & 188 & 0 & 188 & 156 & 93 & 19 & 58 & 41 & 0 & \textbf{367} \\
\midrule
\textbf{Overall Total} & -- & \textbf{228} & \textbf{41} & \textbf{269} 
                       & \textbf{224} & \textbf{111} & \textbf{20} & \textbf{71} & \textbf{59} & \textbf{16} & \textbf{501} \\
\bottomrule
\end{tabular}
\end{table*}

\subsubsection{Evaluation Metrics} 
We evaluated all experiments mainly using three key metrics. Precision quantifies correct positive predictions, Recall measures detection completeness of true positives, and Runtime assesses computational efficiency.

All experiments were conducted on machines equipped with the Intel(R) Core i7-9750H CPU @ 2.60GHz (6 cores and 12 threads) and 16 GB of RAM running 64-bit Ubuntu 22.04 system.

\subsection{RQ1: Effectiveness of \toolname}

To evaluate the effectiveness of \toolname, we analyzed its performance across six trap categories using three LLMs of \toolname: \textit{DeepSeek-v3}, \textit{GPT-5}, and \textit{Gemini-2.5-pro}. Table~\ref{tab:trap_results_comparison} reports precision and recall for each model, along with the average (Avg) values computed as the arithmetic mean across the three LLMs. Notably, we observe a distinct performance trade-off among models: \textit{GPT} achieves the highest overall precision (83.1\%), while \textit{Gemini} excels in recall (89.0\%). \toolname demonstrates robust detection capabilities regardless of the underlying LLM, achieving an average both precision and recall exceeding 0.79 across all LLMs. 


Performance varies significantly across trap categories due to the distinct nature of their underlying logic. For traps rooted in explicit control flow and state dependencies, \toolname achieves high accuracy specifically TR, IM, and TM. For instance, \textit{Tax Manipulation (TM)} attains an average precision of 94.7\%. This success is attributed to the APG's capability to precisely trace data flow from privileged configuration functions (e.g., \texttt{setTax}) to critical execution paths (e.g., \texttt{transfer}), thereby rendering the malicious logic unambiguous to the LLM.
In contrast, while \textit{Balance Tampering (BT)} and \textit{Fake Logs (FL)} achieve exceptional recall (up to 100\%), their precision is notably lower (Avg. 28.2\% for BT). This stems from the semantic ambiguity inherent in advanced DeFi protocols. First, legitimate mechanisms such as rebase tokens (e.g., Ampleforth) or staking rewards often necessitate modifying balances or emitting events outside standard transfer flows, which syntactically resembles trap behavior. Second, to minimize the risk of missed detections (False Negatives), our reasoning module adopts a ``safety-first'' strategy, flagging any non-standard state mutation as a potential threat. While this approach incurs false positives from complex benign logic, it ensures comprehensive coverage of critical assets.

\begin{table}[t]
\centering
\caption{Per-category evaluation results of trap token contracts 
(DeepSeek / GPT / Gemini / Avg)}
\label{tab:trap_results_comparison}
\footnotesize
\begin{tabular}{@{}lc|cccc|cccc@{}}
\toprule
\multirow{2}{*}{Category} & \multirow{2}{*}{\#Samples} 
& \multicolumn{4}{c|}{Precision} 
& \multicolumn{4}{c}{Recall} \\ 
\cmidrule(lr){3-6} \cmidrule(lr){7-10}
& & DeepSeek & GPT & Gemini & Avg 
  & DeepSeek & GPT & Gemini & Avg \\ 
\midrule
TR & 224 
  & 0.923 & \textbf{0.910} & 0.867 & 0.900  
  & 0.906 & \textbf{0.933} & 0.929 & 0.923 \\
IM & 111 
  & \textbf{0.931} & 0.915 & 0.923 & 0.923  
  & 0.784 & \textbf{0.883} & 0.865 & 0.844 \\
IB & 20  
  & 0.684 & 0.665 & \textbf{0.696} & 0.682  
  & 0.550 & 0.600 & \textbf{0.900} & 0.684 \\
TM & 71  
  & 0.919 & \textbf{0.982} & 0.941 & 0.947  
  & 0.634 & 0.761 & \textbf{0.873} & 0.756 \\
BT & 59  
  & \textbf{0.319} & 0.307 & 0.220 & 0.282  
  & 0.695 & 0.796 & \textbf{0.898} & 0.796 \\
FL & 16  
  & 0.469 & \textbf{0.615} & 0.474 & 0.519  
  & 0.938 & \textbf{1.000} & 0.562 & 0.833 \\
\midrule
Overall & 501 
  & 0.829 & \textbf{0.831} & 0.794 & 0.818  
  & 0.803 & 0.870 & \textbf{0.890} & 0.854 \\
\bottomrule
\end{tabular}
\end{table}

\begin{tcolorbox}[
    colframe=black!75, 
    colback=white,     
    boxrule=0.8pt,     
    arc=3pt,           
    left=6pt,          
    right=6pt,         
    top=6pt,           
    bottom=6pt         
]
\textbf{Answer to RQ1:} \toolname is highly effective in detecting diverse trap tokens, achieving an average precision of 0.818 and recall of 0.854. It shows particular strength in identifying logic-based traps (TR, IM, TM) and proves robust in real-world exploit scenarios. While detecting state-ambiguous traps (BT) remains challenging due to false positives from complex DeFi logic, the high recall ensures that potential threats are rarely missed.
\end{tcolorbox}

\begin{table*}[t]
\centering
\caption{Performance comparison on trap detection across different tools}
\label{tab:performance}
\resizebox{\textwidth}{!}{%
\begin{tabular}{l|ccc|ccc|ccc|ccc|ccc}
\toprule
\multirow{2}{*}{\textbf{Category}}
& \multicolumn{3}{c|}{CRPWarner} 
& \multicolumn{3}{c|}{Pied-Piper} 
& \multicolumn{3}{c|}{\toolname (DeepSeek)} 
& \multicolumn{3}{c|}{\toolname (GPT)} 
& \multicolumn{3}{c}{\toolname (Gemini)} \\
\cmidrule(lr){2-16}
& P & R & F1 & P & R & F1 & P & R & F1 & P & R & F1 & P & R & F1 \\
\midrule
TR
& 0.562 & 0.494 & 0.526
& 0.833 & 0.361 & 0.504
& 0.923 & 0.906 & 0.914
& 0.910 & 0.933 & 0.921
& 0.867 & 0.929 & 0.897 \\
IM
& --    & --    & --
& 0.500 & 0.778 & 0.609
& 0.931 & 0.784 & 0.851
& 0.915 & 0.883 & 0.899
& 0.923 & 0.865 & 0.893 \\
IB
& 0.925 & 0.439 & 0.595
& 0.884 & 0.546 & 0.675
& 0.684 & 0.550 & 0.610
& 0.665 & 0.600 & 0.631
& 0.696 & 0.900 & 0.785 \\
TM
& 0.230 & 0.288 & 0.256
& --    & --    & --
& 0.919 & 0.634 & 0.750
& 0.982 & 0.761 & 0.857
& 0.941 & 0.873 & 0.906 \\
BT
& --    & --    & --
& --    & --    & --
& 0.319 & 0.695 & 0.437
& 0.307 & 0.796 & 0.443
& 0.220 & 0.898 & 0.353 \\
FL
& --    & --    & --
& --    & --    & --
& 0.469 & 0.938 & 0.625
& 0.615 & 1.000 & 0.762
& 0.474 & 0.562 & 0.514 \\
Overall
& 0.572    & 0.407    & 0.459
& 0.739    & 0.562    & 0.596
& 0.829 & 0.803 & 0.816
& 0.831 & 0.870 & 0.850
& 0.794 & 0.890 & 0.839 \\
\bottomrule
\end{tabular}%
}
\textsuperscript{*} \footnotesize Dash (--) indicates missing data. Overall is averaged over available categories only.
\end{table*}

\subsection{RQ2: Comparison with SOTA}

We further compared \toolname with two open-source and widely cited tools, \textit{CRPWarner} and \textit{Pied-Piper}. \textit{CRPWarner} identifies contract-related rug pulls via semantic analysis, targeting patterns like hidden minting and token leaking. \textit{Pied-Piper} employs a hybrid approach to expose backdoors, defined as privileged functions for arbitrary asset manipulation (e.g., freezing or burning). 
Unlike these approaches, \toolname combines structured program analysis (ABT+APG) with LLM semantic reasoning; any performance difference thus reflects architectural capability rather than asymmetric knowledge advantage.
To ensure fairness, we aligned their reported vulnerabilities with our taxonomy, following each tool's original detection without modification. The mapping was applied at the result level only: each label was converted to the nearest equivalent category 
, and outputs that could not be cleanly mapped were discarded rather than force-assigned.
Specifically, we mapped mechanisms like ``Hidden Mint'' (CRPWarner) and ``Generate Token'' (Pied-Piper) to \textit{Infinite Mint}, while categorizing their respective locking or freezing features as \textit{Transfer Restrictions}.

CRPWarner and Pied-Piper detect at most three categories, leaving \textit{Tax Manipulation} and \textit{Fake Logs} entirely unaddressed. They completely fail to identify sophisticated logic-based traps, resulting in zero coverage for these categories. \toolname, by contrast, leverages LLMs to interpret the consequences of code execution, achieving comprehensive detection across all six categories.

Even within the categories supported by baselines (TR, IM, IB, TM), \toolname demonstrates superior robustness. Pied-Piper achieves decent precision on \textit{IB} (88.4\%) but low recall (54.6\%). This is because it looks for standard ``burn()'' function calls but misses covert burning mechanisms (e.g., sending tokens to ``address(0)'' via ``transfer''). \toolname captures the semantic equivalent of burning regardless of the implementation syntax, boosting recall to 90.0\% (Gemini). For \textit{TR}, baselines often flag legitimate modifiers as traps (False Positives) or miss restrictions hidden in nested calls (False Negatives). \toolname's APG representation exposes the full dependency chain, allowing the model to distinguish benign governance from malicious locks, yielding an F1-score improvement of almost 0.4 compared to CRPWarner. Overall, \toolname achieves F1-scores between 0.816 and 0.850, significantly outperforming the best baseline (Pied-Piper, F1=0.596).

\begin{table}[t]
\centering
\caption{Runtime decomposition (Average per contract).}
\label{tab:runtime_cost}
\resizebox{\columnwidth}{!}{%
\footnotesize
\begin{tabular}{l|c|c|c}
\toprule
\textbf{LLMs} & \textbf{Preprocessing (ABT+APG) [s]} & \textbf{LLM Inference Time [s]} & \textbf{Total Time [s]} \\
\midrule
\toolname (DeepSeek)  & 23.28 & 128.28 & 151.56 \\
\toolname (GPT)     & 23.28 & 854.92 & 869.20 \\
\toolname (Gemini)    & 23.28 & 891.75 & 915.03 \\
\midrule
Pied-Piper            & --    & --     & 10.17 \\
CRPWarner             & --    & --     & 169.37 \\
\bottomrule
\end{tabular}
}
\end{table}

Table~\ref{tab:runtime_cost} presents the efficiency trade-off. Unsurprisingly, pattern-matching tools like \textit{Pied-Piper} are ultra-fast (10.17s). \toolname incurs higher latency (avg. 151s--915s), primarily driven by the LLM Inference phase. This latency is inherent to our path-based design: each complicated contract generates an average of 43.34 execution paths, and the LLM must reason about each path individually. Among three LLMs, \textit{DeepSeek} stands out for efficiency, completing inference 6 times faster than GPT and Gemini, demonstrating the viability of high-speed reasoning models.

\begin{tcolorbox}[
    colframe=black!75, 
    colback=white,     
    boxrule=0.8pt,     
    arc=3pt,           
    left=6pt,          
    right=6pt,         
    top=6pt,           
    bottom=6pt         
]
\textbf{Answer to RQ2:} \toolname surpasses state-of-the-art static tools by bridging the semantic gap. While baselines are limited to rigid pattern matching (high speed, low coverage), \toolname leverages semantic reasoning to detect complex, obfuscated traps (high coverage, high accuracy).
\end{tcolorbox}

\begin{table}[t]
\centering
\caption{Ablation study of TrapHunter components using the GPT backend. (Precision / Recall / F1).
}
\label{tab:ablation}
\setlength{\tabcolsep}{2.8pt}
\renewcommand{\arraystretch}{1.1}
\begin{tabular}{l ccc ccc ccc ccc}
\toprule
\multirow{2}{*}{\textbf{Category}}
& \multicolumn{3}{c}{\textit{w/o ABT}}
& \multicolumn{3}{c}{\textit{w/o APG}}
& \multicolumn{3}{c}{{\textit{w/o PIA}}}
& \multicolumn{3}{c}{\textbf{TrapHunter (GPT)}} \\
\cmidrule(lr){2-4}\cmidrule(lr){5-7}
\cmidrule(lr){8-10}\cmidrule(lr){11-13}
& P & R & F1
& P & R & F1
& P & R & F1
& P & R & F1 \\
\midrule
TR
& 0.839 & 0.803 & 0.821
& 0.784 & 0.889 & 0.833
& 0.986 & 0.715 & 0.829
& 0.910 & 0.933 & \textbf{0.921} \\
IM
& 0.587 & 0.486 & 0.532
& 0.505 & 0.423 & 0.461
& 1.000 & 0.358 & 0.527
& 0.915 & 0.883 & \textbf{0.899} \\
IB
& 0.255 & 0.264 & 0.259
& 0.214 & 0.170 & 0.189
& 1.000 & 0.353 & 0.522
& 0.665 & 0.600 & \textbf{0.631} \\
TM
& 0.336 & 0.623 & 0.437
& 0.321 & 0.261 & 0.288
& 1.000 & 0.020 & 0.040
& 0.982 & 0.761 & \textbf{0.857} \\
BT
& 0.205 & 0.878 & 0.332
& 0.230 & 0.837 & 0.361
& 1.000 & 0.281 & 0.438
& 0.307 & 0.796 & \textbf{0.443} \\
FL
& 0.081 & 0.750 & 0.145
& 0.080 & 0.875 & 0.147
& 1.000 & 1.000 & 1.000
& 0.615 & 1.000 & \textbf{0.762} \\
\midrule
\textbf{Overall}
& 0.577 & 0.710 & 0.637
& 0.581 & 0.654 & 0.616
& 0.990 & 0.499 & 0.663
& 0.831 & 0.870 & \textbf{0.850} \\
\bottomrule
\end{tabular}
\end{table}

\subsection{RQ3: Ablation Study}
To evaluate the contribution of different components in \toolname, we conducted an ablation study by selectively removing the ABT, APG, and PIA. 
Table~\ref{tab:ablation} presents the per-category results.

Removing ABT results in a comprehensive performance degradation across all categories, with the overall F1-score dropping sharply from 0.850 to 0.637. This universal decline confirms that ABT serves as the foundational layer for code understanding. Specifically, the impact is most illustrative in categories relying on explicit atomic state mutations, such as \textit{IM} and \textit{IB}. For instance, the F1-score for IM plummets by 0.367 (from 0.899 to 0.532). Without ABT, the LLM is forced to parse raw source code, which is often cluttered with ``syntactic noise'' (e.g., complex modifiers, SafeMath wrappers, or assembly blocks). Consequently, the model struggles to pinpoint the critical atomic actions (e.g., \texttt{\_mint} or \texttt{\_balances} updates).

The absence of APG further reduces the overall F1-score to 0.616, proving catastrophic for traps dependent on inter-procedural dependencies. This is most evident in \textit{Tax Manipulation (TM)}. Notably, the F1-score for TM collapses from 0.857 to 0.288, the largest drop among all categories. Tax traps typically involve a \texttt{transfer} function reading a global variable modified by another privileged function. Without the Ternary Traces injected by APG, the LLM views the \texttt{transfer} function in isolation, missing the hidden pathway that weaponizes the logic.

Rather than relying on LLM reasoning, \textit{w/o PIA} directly treats every APG-extracted path as a candidate trap and substitutes a local EVM environment with fuzzing-based transaction generation (exceeding 5M transactions per contract) to probe all six detection predicates exhaustively. As shown in Table~\ref{tab:ablation}, \textit{w/o PIA} achieves near-perfect precision (0.990) but critically low recall (0.499, F1=0.663), confirming that the LLM's primary role is recall recovery: structural matching fails on semantically defined traps, most severely TM (R=0.020). The LLM recovers overall recall from 0.499 to 0.870 by interpreting the meaning of path deviations in context. FL achieves perfect scores (1.000/1.000/1.000) under \textit{w/o PIA}, confirming Fake Logs as a purely structural phenomenon. ABT+APG and LLM-based PIA are complementary: the former surfaces high-precision candidates, the latter recovers recall on semantically complex categories.

\begin{tcolorbox}[
    colframe=black!75, 
    colback=white,     
    boxrule=0.8pt,     
    arc=3pt,           
    left=6pt,          
    right=6pt,         
    top=6pt,           
    bottom=6pt         
]
\textbf{Answer to RQ3:} The three components address orthogonal failure modes. ABT eliminates syntactic noise, with its removal collapsing IM and IB precision to 0.587 and 0.255. APG exposes inter-procedural dependencies invisible within a single function, with its removal causing TM recall to fall from 0.761 to 0.261. PIA provides the semantic interpretation that pure structural matching cannot: without it, overall recall drops to 0.499, as malicious intent encoded in semantics may leave no structural trace. No single component suffices. Their combination is what enables \toolname to achieve both high precision and recall across semantically diverse trap categories.
\end{tcolorbox}

\subsection{RQ4: Validation Analysis}
To ensure that the traps detected by \toolname correspond to triggerable behaviors, we deploy a fork-based validation module to filter out hallucinations identified by the LLM. As shown in Table~\ref{tab:trap_validation}, out of 436 potential traps flagged by the PIA, 373 are successfully replayed and confirmed under forked execution states, yielding a validation rate of 85.6\%. Crucially, for cases where the historical state at the fork block does not satisfy trigger conditions, the validator performs path-constrained state preparation. By executing a limited set of functions strictly identified by the PIA, we ensure that the validation does not introduce arbitrary behaviors or "hallucinate" exploitability.

The validation rate varies across categories due to the complexity of their trigger conditions. Straightforward triggers (IM, IB, FL) rely on direct function invocations with minimal state preconditions, achieving 100\% validation rates. In contrast, complex contextual triggers (TR, TM, BT) often require multi-step state reconstruction, such as a privileged owner first modifying a controlling variable. The lower rates for TR (80.9\%), TM (83.3\%), and BT (70.2\%) are primarily due to cases where these intricate preparatory sequences are not fully reconstructed, leading the validator to conservatively report a failure. Notably, we observed no instances where replayed executions contradicted our formal detection predicates, indicating that the validation stage effectively reduces false positives without introducing new ones.

\begin{table}[t]
\centering
\caption{Detection and validation of trap token contracts (GPT)}
\label{tab:trap_validation}
\footnotesize
\resizebox{\columnwidth}{!}{
\begin{tabular}{l|ccc|ccc|ccc}
\toprule
\multirow{2}{*}{\textbf{Category}}
& \multicolumn{3}{c|}{\#Samples} 
& \multicolumn{3}{c|}{\#Detected Matches (\%)} 
& \multicolumn{3}{c}{\#Validated Paths (\%)} \\
\cmidrule(lr){2-4} \cmidrule(lr){5-7} \cmidrule(lr){8-10}
& ETH-BSC & Backdoor & Total 
& ETH-BSC & Backdoor & Total 
& ETH-BSC & Backdoor & Total \\
\midrule
TR & 68 & 156 & 224 
   & 62 (91.2\%) & 147 (94.2\%) & 209 (93.3\%) 
   & 56 (90.3\%) & 113 (76.9\%) & 169 (80.9\%) \\
IM & 18 & 93 & 111 
   & 18 (100\%) & 80 (86.0\%) & 98 (88.3\%) 
   & 18 (100\%) & 80 (100\%) & 98 (100\%) \\
IB & 1 & 19 & 20 
   & 1 (100\%) & 11 (57.9\%) & 12 (60.0\%) 
   & 1 (100\%) & 11 (100\%) & 12 (100\%) \\
TM & 13 & 58 & 71 
   & 10 (76.9\%) & 44 (75.9\%) & 54 (76.1\%) 
   & 1 (10.0\%) & 44 (100\%) & 45 (83.3\%) \\
BT & 18 & 41 & 59 
   & 18 (100\%) & 29 (70.7\%) & 47 (79.7\%) 
   & 16 (88.9\%) & 17 (58.6\%) & 33 (70.2\%) \\
FL & 16 & 0 & 16 
   & 16 (100\%) & -- & 16 (100\%) 
   & 16 (100\%) & -- & 16 (100\%) \\
\midrule
All & 134 & 367 & 501 
    & \multicolumn{3}{c|}{436 (87.0\% of samples)} 
    & \multicolumn{3}{c}{373 (85.6\% of detected)} \\
\bottomrule
\end{tabular}}
\end{table}


\begin{table}[t]
\centering
\caption{Path redundancy of true-positive (TP) detections and cross-model FN analysis.}
\label{tab:robustness}
\begin{tabular}{l r r r r | r r r r}
\toprule
\multirow{2}{*}{\textbf{Category}}
& \multicolumn{4}{c|}{\textbf{Path Redundancy}}
& \multicolumn{4}{c}{\textbf{False Negatives}} \\
\cmidrule(lr){2-5}\cmidrule(lr){6-9}
& \textbf{TP}
& $\geq$\textbf{2} & $\geq$\textbf{3} & $\geq$\textbf{5}
& \textbf{GPT} & \textbf{Gemini} & \textbf{DeepSeek} & \textbf{All-3} \\
\midrule
TR & 211 & 196 (93\%) & 164 (78\%) & \textbf{117 (55\%)}
   & 15 & 16 & 21 &  4 \\
IM &  96 &  70 (73\%) &  67 (70\%) &  \textbf{63 (66\%)}
   & 13 & 15 & 24 & 11 \\
IB &  12 &   7 (58\%) &   5 (42\%) &   \textbf{4 (33\%)}
   &  8 &  2 &  9 &  2 \\
TM &  54 &  46 (85\%) &  13 (24\%) &   \textbf{6 (11\%)}
   & 17 &  9 & 26 &  7 \\
BT &  47 &  42 (89\%) &  36 (77\%) &  \textbf{17 (36\%)}
   & 12 &  6 & 18 &  3 \\
FL &  16 &   2 (12\%) &   2 (12\%) &   \textbf{1 (~6\%)}
   &  0 &  7 &  1 &  0 \\
\midrule
\textbf{Total} & \textbf{436}
& \textbf{363} & \textbf{287} & \textbf{208}
& \textbf{65} & \textbf{55} & \textbf{99} & \textbf{27} \\
\bottomrule
\end{tabular}
\end{table}

\textbf{FP Analysis.} Post-hoc inspection of all 63 rejected cases (436 detected$-$373 validated) confirms they are genuine LLM FPs---none were actual traps incorrectly rejected by the validator. This confirms that the formal predicates in Table~\ref{tab:formal_trap_table} serve as a reliable execution-grounded filter, achieving zero validator-induced FPs. The 63 FPs concentrate in three categories: TR (40), BT (14), and TM (9), each with a distinct structural origin.
\textbf{(i) TR:} The APG detects an external function that can modify a variable along the transfer path; however, the variable governs balance calculations rather than transfer access control. The fork validation finds no unauthorized control on transfer execution, correctly rejecting these 40 cases.
\textbf{(ii) BT:} Legitimate DeFi mechanisms directly modify \texttt{balances[]} via privileged functions, producing execution paths structurally identical to BT traps. The LLM accurately detects the balance modification but cannot differentiate authorized administrative intent from unauthorized tampering without broader contract contexts. Fork validation bounds these FPs: only 33 of 47 BT-flagged contracts are confirmed.
\textbf{(iii) TM:} Contracts with complex but authorized fee structures (e.g., protocol fees) satisfy TM detection criteria. Fork execution confirms that fee changes remain within authorized bounds, correctly rejecting these 9 cases as FPs.

\textbf{FN Analysis.} TrapHunter's FN risk arises from three independent sources. \textbf{(i) APG-induced FN}: ABT+APG may fail to generate any path through the trap function due to complex inheritance. ABT mitigates this by resolving the full contract hierarchy and systematically tracing the inheritance chain. Among 65 FN samples, only 20 (30.8\%) exhibit zero APG coverage, confirming that reasoning failure rather than coverage gaps is the dominant FN cause. 
\textbf{(ii) LLM-induced FN}: PIA may miss traps due to reasoning limitations even when the relevant path exists, mitigated by majority voting within each model. Beyond single-model mitigation, we evaluate \toolname across three LLMs (DeepSeek, GPT, Gemini) in Table~\ref{tab:robustness} and treat a contract as a persistent FN only if all three models miss it (All-3 column). This cross-model consensus reduces total FN from 65 to 27 (58.5\%), confirming that the FNs can be corrected by model ensemble or substitution.
\textbf{(iii) Triggering-induced FN}: fork validation may fail to reconstruct required on-chain states for dormant traps; path-constrained state preparation mitigates this, but cannot guarantee full coverage.

\textbf{Detection Robustness.} In Table~\ref{tab:robustness}, the right section decomposes FNs across three LLM backbones, while the left section quantifies path redundancy among confirmed TP contracts.  A contract is flagged if any path is confirmed as a trap, so detection does not rely on a single path being correct. Overall, 363 of 436 TP contracts (83.3\%) have at least two independent APG paths simultaneously reaching the same verdict.  For four of six categories, this rate exceeds 70\%: TR (93\%), BT (89\%), TM (85\%), and IM (73\%). IB reaches 58\%, still indicating that the majority of its TP contracts are multiply confirmed.  Even when individual path-level judgments are uncertain, the contract-level verdict remains stable through multi-path consensus.  The sole exception is FL (12\%), where trap behavior manifests on a single structural path, directly explaining its lower path redundancy.

\begin{tcolorbox}[
    colframe=black!75, 
    colback=white,     
    boxrule=0.8pt,     
    arc=3pt,           
    left=6pt,          
    right=6pt,         
    top=6pt,           
    bottom=6pt         
]
\textbf{Answer to RQ4:} The fork-based validation module acts as a vital reliability layer. It confirms 85.6\% of detected traps as practically exploitable while effectively filtering out 14.4\% of cases, ensuring high-confidence reporting. 
\end{tcolorbox}

\subsection{RQ5: Generalization to Real-World On-Chain Contracts}
\label{sec:onchain-dataset}

The on-chain dataset was constructed by scanning the Ethereum Mainnet covered blocks 24,949,685-24,955,690 ($\approx 6,000$ blocks), yielding 500 unique candidate contract addresses after deduplication. These candidates were then sequentially queried via the Etherscan API to retrieve verified Solidity source code. The retrieval process halted once 50 contracts with verifiable source code were collected. Two were excluded during preprocessing (1 ERC721 contract, 1 unparsable JSON bundle). After preprocessing, the evaluation set was finalized at \textbf{48 contracts} containing 1,563 execution paths. Ground truth (\#Contracts)  was established through two-phase manual inspection by two authors, with disagreements resolved by a third senior researcher.

Table~\ref{tab:result_new} reports the detection results. \toolname achieves an overall precision of 0.806 and a perfect recall of 1.000 (F1=0.893). Among specific categories, TR demonstrates robust performance with an F1 of 0.927, confirming that logic-based traps remain identifiable across diverse real-world deployments. FL and TM exhibit lower precision (0.077 and 0.545). For FL, 12 out of 13 detections are FPs caused by legitimate fees on transfer tokens, indicating that distinguishing deceptive logs from complex fee-taxing logic remains a challenge. Similarly, TM's FPs primarily arise from non-malicious administrative fee adjustments. The absence of detected BT cases reflects its limited frequency in real-world samples, rather than the non-existence of this trap pattern. Notably, the overall recall of 1.000 is attributable to the fact that all malicious contracts in the dataset exhibited multiple trap patterns, and \toolname successfully flagged at least one in every contract.

\begin{table}[t]
\centering
\caption{Detection results of \toolname (GPT) on the on-chain Ethereum dataset (48 contracts).}
\label{tab:result_new}
\begin{tabular}{l c cccccc c}
\toprule
& \textbf{Ground truth} & \multicolumn{6}{c}{\textbf{Detection}} & \textbf{Verification} \\
\cmidrule(lr){2-2} \cmidrule(lr){3-8} \cmidrule(lr){9-9}
\textbf{Trap} & \textbf{\#Contracts} & \textbf{TP} & \textbf{FP} & \textbf{FN} & \textbf{P} & \textbf{R} & \textbf{F1} & \textbf{Val.\,(\%)} \\
\midrule
TR & 20 & 19 & 2  & 1 & 0.905 & 0.950 & 0.927 & 85.0 \\
IM & 6  & 4  & 0  & 2 & 1.000 & 0.667 & 0.800 & 66.7 \\
IB & 5  & 5  & 2  & 0 & 0.714 & 1.000 & 0.833 & 80.0 \\
TM & 6  & 6  & 5  & 0 & 0.545 & 1.000 & 0.706 & 83.3 \\
BT & 0  & 0  & 0  & 0 & --    & --    & --    & --   \\
FL & 1  & 1  & 12 & 0 & 0.077 & 1.000 & 0.143 & 100.0 \\
\midrule
\textbf{Overall} & \textbf{25} & \textbf{25} & \textbf{6} & \textbf{0} & \textbf{0.806} & \textbf{1.000} & \textbf{0.893} & \textbf{--} \\
\bottomrule
\end{tabular}
\end{table}

\begin{tcolorbox}[
    colframe=black!75,
    colback=white,
    boxrule=0.8pt,
    arc=3pt,
    left=6pt, right=6pt, top=6pt, bottom=6pt
]
\textbf{Answer to RQ5:} \toolname generalizes effectively to live environments, achieving an overall F1 of 0.893 (R=1.000) across 48 freshly collected Ethereum Mainnet contracts. Its high precision in critical categories such as TR (0.905) and IM (1.000) underscores its reliability in identifying the most high-impact trap patterns in the wild.
\end{tcolorbox}

\subsection{Case Study}
To demonstrate how \toolname uncovers hidden dependencies, we conduct a case study on the \texttt{BitDao} contract. Unlike traditional static analysis tools that view functions in isolation, \toolname's APG constructs a holistic view by injecting state-coupled functions into the execution path. Table~\ref{tab:case_study} details four representative paths extracted by our tool. 

\begin{table*}[t]
\centering
\caption{Representative Augmented Paths and Trap Detection Analysis on \texttt{BitDAO}.}
\label{tab:case_study}
\footnotesize 
\begin{tabularx}{\textwidth}{l p{3.7cm} p{4cm} X l}
\toprule
\textbf{ID} & \textbf{Augmented Execution Path (Simplified)} & \textbf{Detected Dependency (Ternary Trace)} & \textbf{Trap Logic} & \textbf{Verdict} \\
\midrule
\textbf{P1} & 
$\top \rightarrow$ transfer $\rightarrow$ \_approveCheck $\rightarrow$ \_beforeTokenTransfer $\rightarrow$ \_msgSender $\rightarrow \bot$ & 
\texttt{[burnTokenCheck(Modifier) \textbf{READS} \_safeOwner \textbf{WRITTEN\_BY} decreaseAllowance(public)]} & 
Modifier Ternary Traces - Public function toggles transfer lock. & 
TR \\
\midrule
\textbf{P2} &
$\top \to$ \texttt{multiTransfer} $\to$ \texttt{\_approve} $\to \bot$ &
\texttt{[multiTransfer(public) \textbf{READS} \_whiteAddress \textbf{WRITTEN\_BY} increaseAllowance(public)]} & 
Function Ternary Traces - Public function bypasses permission check. & 
TR \\
\midrule
\textbf{P3} &
$\top \to$ \texttt{\_mint} $\to \bot$ &
\texttt{[\_mint(public) \textbf{READS} \_totalSupply]} & 
Uncapped minting by public owner. & 
IM \\
\midrule
\textbf{P4} &
$\top \rightarrow \_burn \rightarrow \_beforeTokenTransfer \rightarrow \bot$ &
\texttt{[\_burn(internal) \textbf{READS} \_totalSupply]} & 
Internal function, no public entry. & 
No Trap \\
\bottomrule
\end{tabularx}
\end{table*}

Path P1 \& P2 illustrate the importance of the \textbf{Ternary Traces} construction rules.
Path P1 exposes a sophisticated \textit{Transfer Restriction (TR)}. A standard call graph would only show \texttt{transfer} invoking the modifier \texttt{burnTokenCheck}. Since the modifier's logic appears benign (checking a boolean), superficial scanners ignore it. However, \toolname's APG injects a Modifier Ternary Trace, revealing a hidden coupling: \texttt{burnTokenCheck} reads the variable \texttt{\_safeOwner}, which can be arbitrarily toggled by a completely separate public function, \texttt{decreaseAllowance}. By linking the \textit{victim's action} (\texttt{transfer}) with the \textit{attacker's switch} (\texttt{decreaseAllowance}), \toolname successfully identifies this as a TR trap with lock switch. Similarly, in Path P2, the system detects that the whitelist variable \texttt{\_whiteAddress} is manipulable via \texttt{increaseAllowance}, flagging a backdoor that allows attackers to bypass permissions. 

Finally, we compare Path P3 and P4 to illustrate how the \texttt{Root} layer of \toolname exposes actual threats. Path P3 is correctly flagged as an \textit{Infinite Mint (IM)} trap because the APG detects a direct state mutation on \texttt{\_totalSupply} within a publicly accessible function (\texttt{\_mint}). Conversely, Path P4 demonstrates the system's ability to filter false positives. Although the \texttt{\_burn} function technically contains logic to destroy unlimited tokens (resembling an Infinite Burn trap), our analysis identifies that its ``Writer Context'' is strictly \texttt{internal} with no public reachability graph connecting to it. Consequently, \toolname correctly classifies P4 as a benign internal utility, avoiding the false alarms common in pattern-matching tools.

\begin{tcolorbox}[
    colframe=black!75, 
    colback=white,     
    boxrule=0.8pt,     
    arc=3pt,           
    left=6pt,          
    right=6pt,         
    top=6pt,           
    bottom=6pt         
]
\textbf{Summary:} This case study confirms that \toolname's path-based analysis goes beyond syntax. By explicit modeling Ternary Traces (P1/P2) and ABT (P3 vs. P4), it successfully connects disparate code components to expose traps hidden in code, while correctly ignoring unreachable internal logic to ensure high precision.
\end{tcolorbox}

\subsection{Threats to Validity}

Our study has several potential threats. \textbf{First}, LLM may yield non-deterministic results, mitigated by temperature-varied majority voting (Section~4.4) and confirmed robust through three-model ensemble analysis (Table~\ref{tab:robustness}). \textbf{Second}, sample bias may exist; the De.Fi REKT database mitigates this by ensuring real-world coverage rather than theoretical examples, and the on-chain evaluation (Section~5.6) further validates \toolname on unseen deployed contracts.
\textbf{Third}, triggering-induced FN may occur if required on-chain states cannot be reconstructed for dormant traps, which path-constrained state preparation mitigates but cannot fully guarantee (Section~5.5). \textbf{Fourth}, APG path coverage may be incomplete if adversarially obfuscated call chains evade our inheritance-aware path generation, constituting a structural FN independent of LLM reasoning quality (Section~5.5).

\section{Related Work}

\subsection{Traditional Static and Symbolic Analysis}
Early detection~\cite{wang2020oracle, wang2019vultron} primarily relied on heuristic rules and symbolic execution. 
\textsc{HoneyBadger}~\cite{torres2019art} and \textsc{HoneyToken-Detector}~\cite{liu2023honeytoken} utilize symbolic execution to match control flow paths against manually crafted predicates for known honeypots. 
General-purpose analyzers like \textit{Slither} and \textit{Mythril} target code vulnerabilities (e.g., reentrancy) but lack rules for logic-based traps. 
Formal verification tools like \textsc{VerX}~\cite{permenev2020verx} and \textsc{K-Framework}~\cite{hildenbrandt2018kevm} provide mathematical guarantees but face scalability bottlenecks with complex DeFi inheritance. 
Specification mining further automates property inference from execution traces~\cite{liu2025specification}.
More recently, \textsc{ZepScope}~\cite{liu2024using} investigated scams specifically exploiting OpenZeppelin libraries.

\subsection{Learning-based Fraud Detection}
To overcome the rigidity of rules, researchers applied machine learning (ML) and graph analysis.
\textbf{Feature-based ML:} Early works employed N-gram features~\cite{chen2020honeypot}, opcode sequences~\cite{hara2021machine}, or multi-modal features~\cite{camino2020data} with traditional classifiers.
Deep learning models such as \textsc{SCSGuard}~\cite{hu2022scsguard}, \textsc{DeFiTrust}~\cite{gunathilaka2024defitrust}, and others further improved detection~\cite{gu2023trap} by learning latent patterns from bytecode. DeFiScanner~\cite{wang2022defiscanner}, DEFIER~\cite{su2021evil}, MoTS~\cite{wu2023know}, and MetaSuites~\cite{MetaSuites}, utilize program features but still suffer from some limitations in obtaining potential malicious semantic features. TXR-TCC~\cite{wu2025hunting} train models with attack/non-attack event features, leading to high resource costs. 
\textbf{Graph-based Detection:} Recognizing the importance of transaction context, tools like \cite{chen2025ethereum} extract higher-order semantics and apply heterogeneous graph transformers for classification, but require large-scale training data.
Graph features (e.g., \cite{wu2024tokenscout}) often rely on post-attack transaction logs, rendering them unsuitable for early-stage preventive detection at the code level.

\subsection{LLM-Driven Smart Contract Security}
Large Language Models (LLMs) have demonstrated remarkable capabilities in code understanding and reasoning. Recent frameworks like \textsc{TrustLLM}~\cite{ma2024combining} and \textsc{PropertyGPT}~\cite{liu2024propertygpt} utilize LLMs for automated auditing and formal property generation, identifying logic flaws that elude traditional tools~\cite{sun2024gptscan}. 
DeepTx~\cite{liu2025deeptx} combines multi-modal transaction features with LLM reasoning for real-time transaction risk analysis.
Furthermore, research on automated repair, such as \textsc{ACFix}~\cite{zhang2025acf} and \textsc{VulAdvisor}~\cite{zhang2024vuladvisor}, highlights the importance of guiding LLMs with mined patterns or local context to fix complex access control vulnerabilities.
Augmenting prompts with semantic facts significantly boosts LLM performance over raw code~\cite{ahmed2024automatic}, which motivates our ABT/APG-based prompt design.

\section{Conclusion}
We proposed \toolname, an end-to-end intent deviation detection framework for trap tokens. \toolname constructs a unified semantic representation using ABTs and APGs to normalize intra-procedural syntax and expose inter-procedural state dependencies, then applies LLM-powered reasoning to distinguish malicious intent, and finally confirms exploitability via fork-based dynamic validation.
Evaluation on 269 real-world contracts (501 labeled samples) achieves an average precision of 81.8\% and recall of 85.4\%, significantly outperforming state-of-the-art tools, with generalization to 48 unseen on-chain contracts confirming robustness beyond curated benchmarks.

\section{Data Availability}
Our replication package is available online: 
\href{https://doi.org/10.6084/m9.figshare.30082903}{\texttt{https://doi.org/10.6084/m9.figshare.30082903}}


\begin{acks}
This research was supported by National Natural Science Foundation of China (62372367, 62232014, 62272377, 62372368), Shaanxi Province Sanqin Talent Introduction Program, the Singapore Ministry of Education Academic Research Fund Tier 2 (T2EP20224-0003) and the Nanyang Technological University Centre for Computational Technologies in Finance (NTU-CCTF). Any opinions, findings, and conclusions or recommendations expressed in this material are those of the author(s) and do not necessarily reflect the views of MOE and NTU-CCTF.
The first author acknowledges the financial support from the China Scholarship Council.

\end{acks}

\bibliographystyle{ACM-Reference-Format}
\bibliography{sample-base}




\end{document}